\documentclass[8.5pt,twoside,twocolumn]{article}
\usepackage[super,sort&compress,comma]{natbib} 
\usepackage{mhchem}
\usepackage{balance}
\usepackage{times,mathptm}
\usepackage{sectsty}
\usepackage{graphicx} 
\usepackage{lastpage}
\usepackage[format=plain,singlelinecheck=false,font=small,labelfont=bf,labelsep=space]{caption} 
\usepackage{fancyhdr}
\begin{document}

\twocolumn[
  \begin{@twocolumnfalse}
\noindent\LARGE{\textbf{Strain Stiffening Induced by Molecular Motors in Active Crosslinked Biopolymer Networks}}
\vspace{0.6cm}

\noindent\large{\textbf{Peng Chen\textit{$^{a}$} and Vivek B. Shenoy*\textit{$^{a}$}}}\vspace{0.5cm}

 \end{@twocolumnfalse} \vspace{0.6cm}

  ]

\noindent\textbf{We have studied the elastic response of actin networks with both compliant and rigid crosslinks by modeling molecular motors as force dipoles. Our finite element simulations show that for compliant crosslinkers such as filamin A, the network can be stiffened by two orders of magnitude while stiffening achieved with incompliant linkers such as scruin is significantly smaller, typically a factor of two, in excellent agreement with recent experiments. We show that the differences arise from the fact that the motors are able to stretch the compliant crosslinks to the fullest possible extent, which in turn causes to the deformation of the filaments. With increasing applied strain, the filaments further deform leading to a stiffened elastic response. When the crosslinks are incompliant, the contractile forces due to motors do not alter the network morphology in a significant manner and hence only small stiffening is observed.}
\section*{}
\vspace{-1cm}

\footnotetext{*\textit{$^{a}$~School of Engineering, Brown University, Providence, RI 02912, USA. Email:Vivek\_Shenoy@brown.edu}}


The mechanical properties of plant and animal cells are governed by the cytoskeleton, a flexible and dynamic network of biopolymer fibers combined with a group of associated regulatory and crosslinking proteins~\cite{Howard01, Fletcher10}. One of the key aspects of the mechanical behaviour of these networks is their highly nonlinear elastic response to applied stresses~\cite{Gardel06}, in particular their ability to strain stiffen by orders of magnitude when subject to large stresses.  Cells also employ molecular motors to convert chemical energy into mechanical work~\cite{Howard01}. Motors generate internal stress in the networks even in the absence of external loading~\cite{Howard01, Mizuno07}. In this manner, cells can regulate their mechanical properties by using both active and passive components.

While the mechanical behavior of semiflexible polymer networks with compliant and rigid crosslinks has been studied in detail both experimentally and theoretically~\cite{Onck05, Brodersz08, Koenderink09, Kasza09, Lin10, Lieleg10}, the interplay between active mechanisms of stress generation  through motor activity and passive strain hardening properties of crosslinks has only been considered very recently~\cite{Mizuno07, Koenderink09}. In this regard, reconstituted actin networks can be particularly useful, since the density of  crosslinks and motors can be varied in a desired manner to gain insights into the mechanisms of strain hardening and nonlinear elastic response. Indeed, recent experiments on networks that consist of actin filaments crosslinked by filamin A (FLNa) and bipolar filaments of muscle myosin II show that in the absence of any applied loads~\cite{Koenderink09}, the motors stiffen the network by about two orders of magnitude. The degree of stiffening was found to increase with increasing density of myosin motors. Another key observation from this study relates to the magnitude of stiffening caused by compliant and incompliant crosslinks. While FLNa is a large, highly flexible dimer that promotes orthogonal F-actin crosslinking, scruin is an incompliant crosslink. Interestingly, it was found that in distinct contrast to FLNa, scruin does not promote active stiffening of F-actin networks upon addition of myosin. These results clearly show that actomyosin contractility when combined with appropriate crosslinks can allow the cell to operate in a nonlinear regime to actively control its mechanical response.

\begin{figure}[h!]
\centering
\includegraphics[height=2.4in]{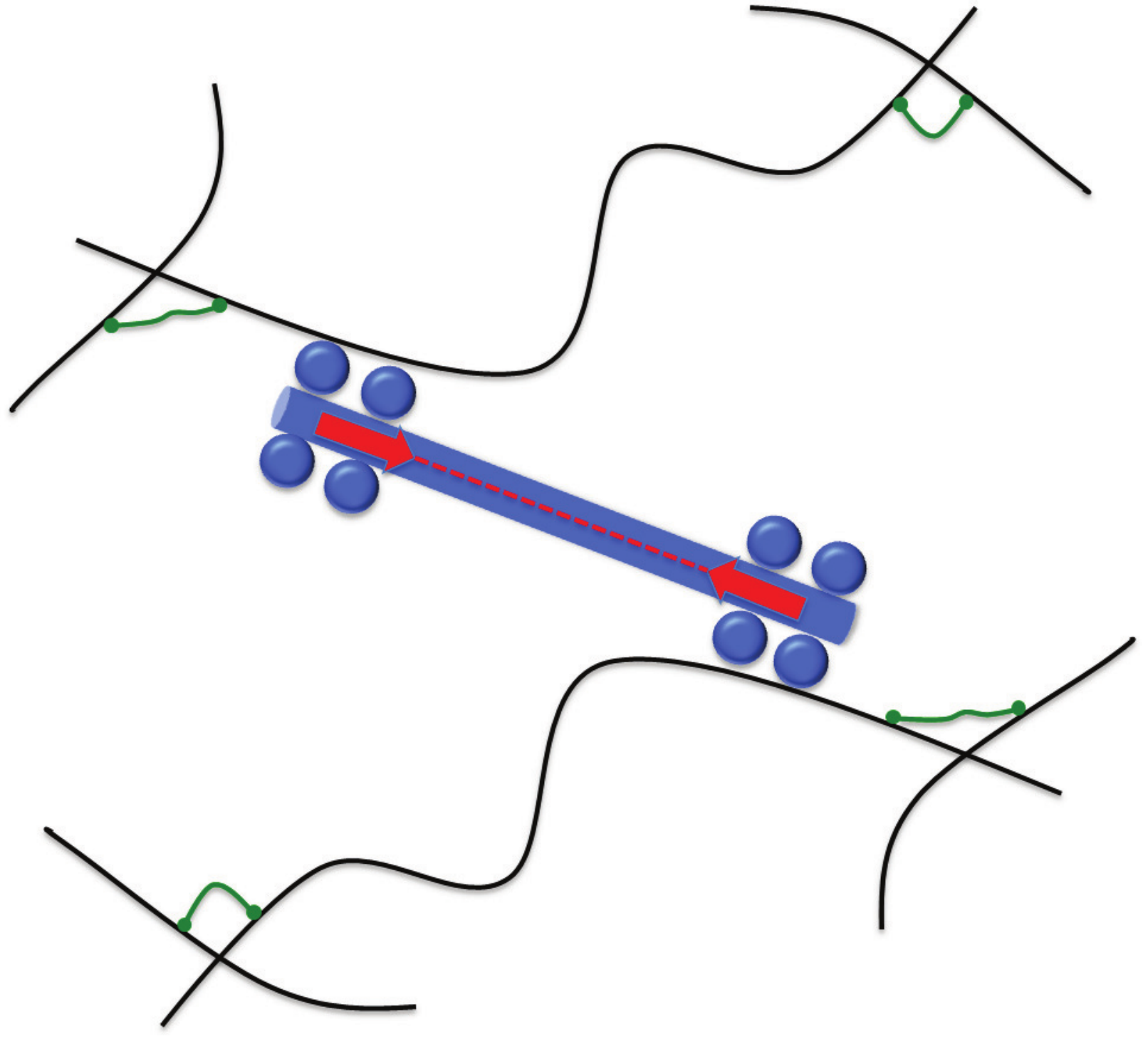}       
 \caption{Schematic of actin filaments (black lines) with myosin motors (blue) and compliant crosslinks (green lines). Myosin motor exerts equal and opposite forces on the filaments on which it is attached, which results in a force dipole (pair of red arrows). These forces lead to the extension of the filamin crosslinks.}
  \label{fig:motor}
\end{figure}

Why do compliant crosslinks in active networks lead to large strain stiffening while no significant increase in stiffness is observed in the case of incompliant crosslinks? To answer this question and to quantitatively study the interplay of internal strains generated by molecular motors and external loads, we study the mechanical response of active networks using finite element simulations where the motors are treated as force dipoles (Fig.~\ref{fig:motor}). A number of approaches including mean-field models, effective medium theory and numerical simulations have been used to study the elastic response of passive networks with both compliant and incompliant crosslinks~\cite{Onck05, Brodersz08}. However the role of crosslinks on the mechanical response of networks with molecular motors has not been considered theoretically. Our work shows that the nature and density of the crosslinks play a key role in determining the strain stiffening response in active biopolymer networks. FLNa is a crosslink that is compliant at small pulling forces, but is stiff beyond a critical value of stretch~\cite{Furuike01}. We find that even in the absence of applied loads, motors lead to almost completely stretch out the compliant crosslinks taking them into the stiffened regime, which in turn also leads to bending of the filaments. As the crosslinks are fully stretched, the applied load is accommodated by the deformation of filaments. Since it is more difficult to deform the filaments compared to stretching of compliant crosslinks, the network stiffens by as much as two orders of magnitude compared to the case when motors are absent. On the other hand, when the crosslinks are incompliant, the contractile forces due to motors do not alter the network morphology in a significant manner leading to much lesser stiffening (typically by a factor of two) in agreement with experiments~\cite{Koenderink09}.

The biopolymer networks considered in our study are constructed as follows. Straight filaments of  length $L$ = 10 $\mu$m (comparable to the actin filament lengths of $\sim$10-15 $\mu$m in Ref.\cite{Koenderink09}) and random orientations are assembled in a square box of width  $W =$ 40 $\mu$m. When two filaments cross each other, they are connected by a nonlinear spring (to be described below) or are linked rigidly. Filaments that cross the top and bottom boundary are cut and the dangling ends are removed, while periodic boundary conditions are imposed along the lateral  boundaries. The system is loaded by restricting the horizontal displacements of the nodes of the filaments at the top and bottom of the box to  $\gamma W$ and zero, respectively, where $\gamma$ is the applied shear strain. We assume that the polymer filaments that make up our network are semiflexible, so that the persistence length of the individual chains is much longer than the average distance between the two crosslink sites, and comparable to the contour length of individual polymer chains. Therefore we ignore thermal energy arising from fluctuation of the filaments and consider only the extensional and bending energies of the filaments. Using $\kappa$ and $\mu$ to denote bending modulus and stretching modulus of F-actin, respectively, we choose  $\mu /L = 1.6$ MPa  and $\sqrt{\kappa/(\mu L^2)} = 2.3 \times 10^{-4}$ (both representative of F-actin networks \cite{Howard01,Onck05}). The normalized line density of our network ($\bar{\rho} = L_TL/W^2$, where $L_T$ is the total length of filaments in the  cell) is $\bar{\rho} = 12.5$, which is well above the rigidity percolation threshold of $\bar{\rho} = 5.7$ \cite{Wilhelm03}. 

The contour length of FLNa is $l_0 = 150$ nm and its persistence length is $l_p$ = 20 nm \cite{Furuike01}.  For stretches less than the contour length, this crosslink behaves like a linear spring with spring constant  $k_{cl} = \frac{2 k_B T}{3 l_p l_0}$  \cite{Marko94}, where $k_B T = 4.11$ pN $\cdot$ nm.  When stretched beyond its contour length $l_0$, the stiffness of the crosslink increases very rapidly\cite{Furuike01, Brodersz08}. Following the work of Brodersz et al.~\cite{Brodersz08}, we therefore model the crosslinks as piecewise linear springs such that the force $F = -k \Delta l$, where $k = k_{cl}$ for length $\Delta l < l_0$, while $k \gg k_{cl}$ when $\Delta l > l_0$, where $\Delta l$ is the extension of the crosslink. Scruin crosslinks on the other hand are inextensible~\cite{Gardel06}. Very recent measurement shows that force applied by a single skeletal myosin head on an F-actin is in the range 1pN - 5pN~\cite{Kaya10}. Myosin is typically assembled into thick filaments with several heads; the filaments typically have lengths of $\sim 1\mu$m \cite{Svitkina95, Koenderink09}. To model the contractile force generated by the motors we model them as force dipoles as shown schematically in Fig.~\ref{fig:motor}. In our simulations, the motors are assembled by picking a point at random on a filament and then picking another point located on a neighboring filament such that the distance of the two points lies in the range of 0.5-2 $\mu$m. A typical force of 10pN is applied along the line connecting the two points to simulate the pulling effect of the motors as shown in Fig.~\ref{fig:motor}. 
\begin{figure}[h!]
 \centering
\includegraphics[height=3in]{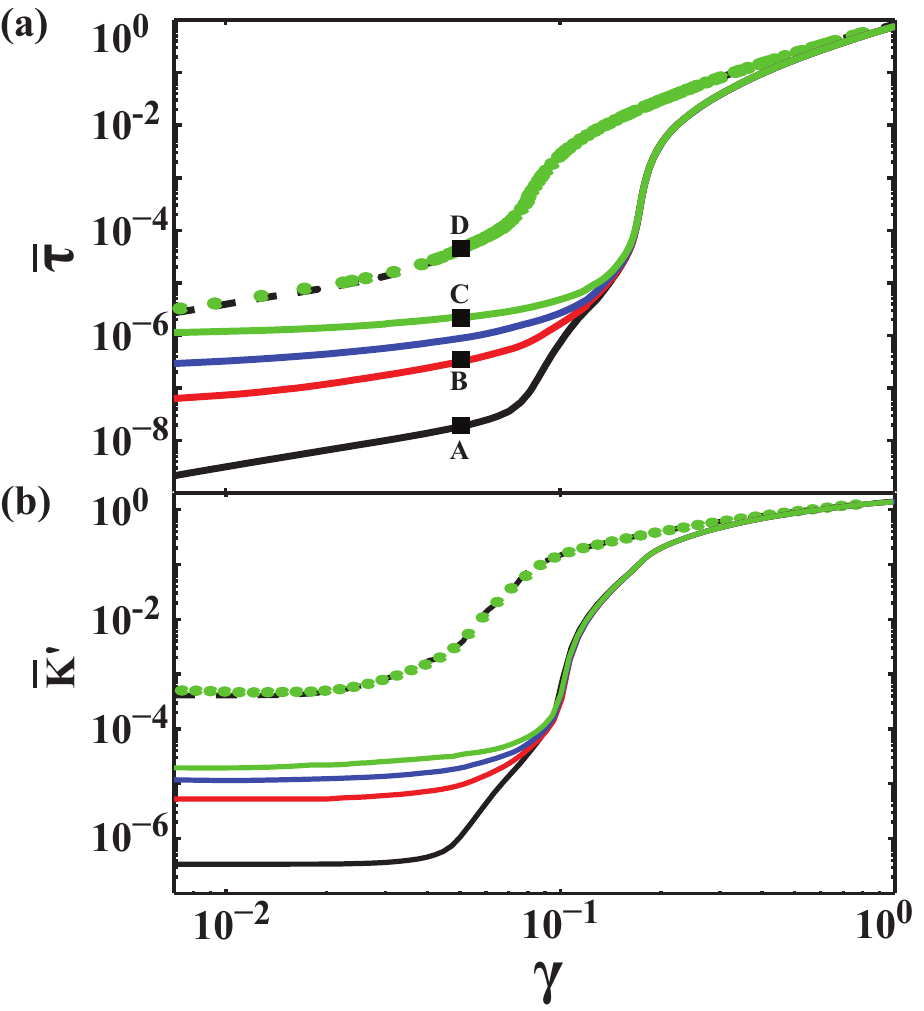}       
  \caption{Stress ($\bar{\tau}$) and differential shear stiffness ($\overline{K'}$) versus shear strain ($\gamma$) for active F-actin network with rigid or compliant crosslinks. The myosin motors only stiffen the F-actin network with rigid crosslinks by a factor of 1.4 at small strains: network without myosin motor (black dashed line) and $N_m/N_{actin} = 6.4$ (green circles). However the myosin motors stiffen the flexibly crosslinked F-actin network up to a factor of 70: network without myosin motors (black solid line), with $N_m/N_{actin} = 0.9$ (red solid line), with $N_m/N_{actin} = 3.1$ (blue solid line), and with $N_m/N_{actin} = 6.4$ (green solid line). $N_m$ and $N_{actin}$ correspond to the total number of motors and filaments, respectively. Each motor exerts a force of 10pN on the actin filament.}
  \label{fig:stiff}
\end{figure}

\begin{figure}[h!]
 \centering
 \includegraphics[height=3in]{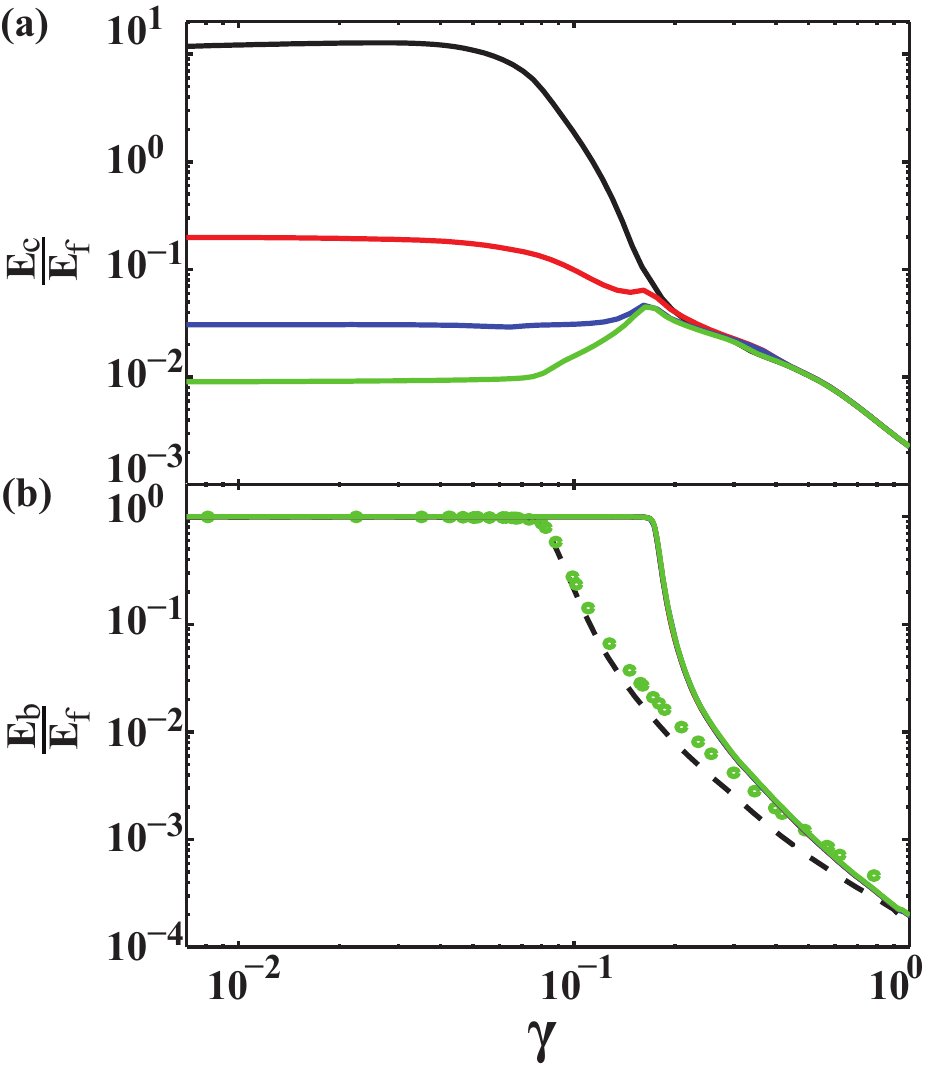}       
  \caption{(a) The ratio of total stretching energy ($E_c$) of compliant crosslinks to total energy ($E_f$) of filaments in the system, and (b) the ratio of total bending energy ($E_b$) of the filaments in the system to total strain energy of all the filaments ($E_f$) as a function of shear strain ($\gamma$). At small strains, the rigidly crosslinked F-actin network is dominated by bending of F-actin, regardless of whether myosin motors are present (green circles) or absent (black dashed line). However, myosin motors drive the deformation from stretching of the crosslinks to bending of F-actin as evidenced by the decrease in the ratio $E_c/E_f$ from the case with no motors  (black solid line), to the cases with increasing contractile forces (not plotted here) or motor densities ($N_m/N_{actin} = 0.9$ (red solid line), $N_m/N_{actin} = 3.1$ (blue solid line), and $N_m/N_{actin} = 6.4$ (green solid line)). All flexibly crosslinked F-actin networks with or without myosin motors show nearly identical behavior of the ratio $E_b/E_f$ (green solid line in (b)).}
  \label{fig:energy}
\end{figure}

The elastic fields in the sheared filament network are computed using the finite element method, discretizing each filament with 100 equal-sized Timoshenko beam elements~\cite{Timoshenko}. All simulations are carried out in a finite deformation setting; i.e., the effect of geometry changes on force balance and rigid body rotations are explicitly taken into account. The macroscopic shear stress $\tau$ for our model is the total horizontal reaction force at the top of the simulation box divided by $W$. The dimensionless stress $\bar{\tau}$ in Fig.~\ref{fig:stiff} is defined as $\bar{\tau} = \tau L/\mu$, and the modulus $\overline{K'} = d\bar{\tau}/d\gamma$. To gain insight into different deformation mechanisms, we also compute the total energy associated with stretching of the crosslinks in the cell, $E_c$, the total bending energy of all the filaments, $E_b$, and the sum of the bending and stretching energies or the total strain energy of all the filaments, $E_f$. We first consider the case where there are no motors present and the crosslinks are either rigid (black dashed lines in Figs.~\ref{fig:stiff} and~\ref{fig:energy}) or compliant (black solid lines in Figs.~\ref{fig:stiff} and~\ref{fig:energy}). As in earlier work \cite{Brodersz08}, the modulus in the latter case shows a sharp increase when the crosslinks are fully stretched and enter the hardened regime. It can be seen from Fig.~\ref{fig:energy} that below this threshold the total energy of the network is dominated by the energy of stretching the crosslinks, but above this threshold the filaments first bend and then stretch. In the case of rigid crosslinks, for small strains, the deformation of the network is dominated by bending of the filaments followed by stretching and orientation of the filaments along the direction of shearing. Note that at very large applied strains, both the networks show identical response due to stretching of the filaments.

\begin{figure}[h!]
 \centering
 \includegraphics[width=3in]{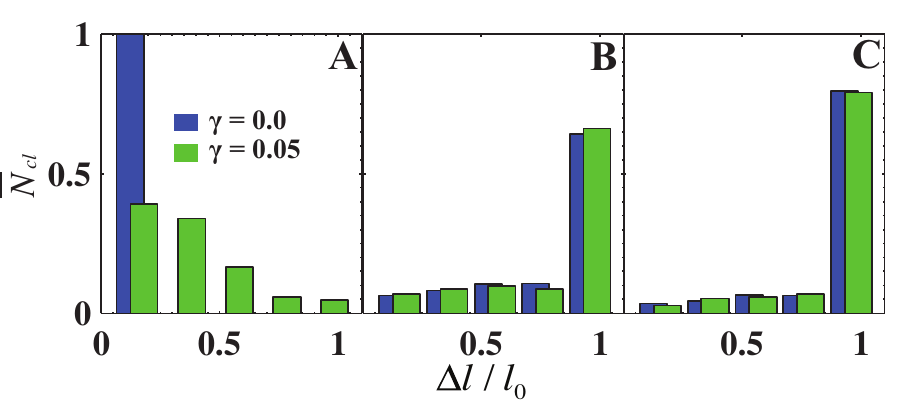}       
  \caption{Distribution of the extension of the crosslinks $\Delta l$ relative to their contour length $l_0$ at shear strains $\gamma = 0$ (blue) and $\gamma = 0.05$ (green). $\bar{N}_{cl}$ is the number of crosslinks normalized by the total number of crosslinks in the active network. The symbols A-C corresponded to the black squares given in Fig.~\ref{fig:stiff}.  In B and C myosin motors drive most compliant crosslinks up to contour lengths, $l_0$,  whereas when no motors are present (A), the crosslinks have not been stretched to their fullest extent. Note that more crosslinks are stretched to their contour lengths in C than in B owing to the larger motor density in the former case.}
  \label{fig:clhist}
\end{figure}

\begin{figure}[h]
 \centering
 \includegraphics[width=3.3in]{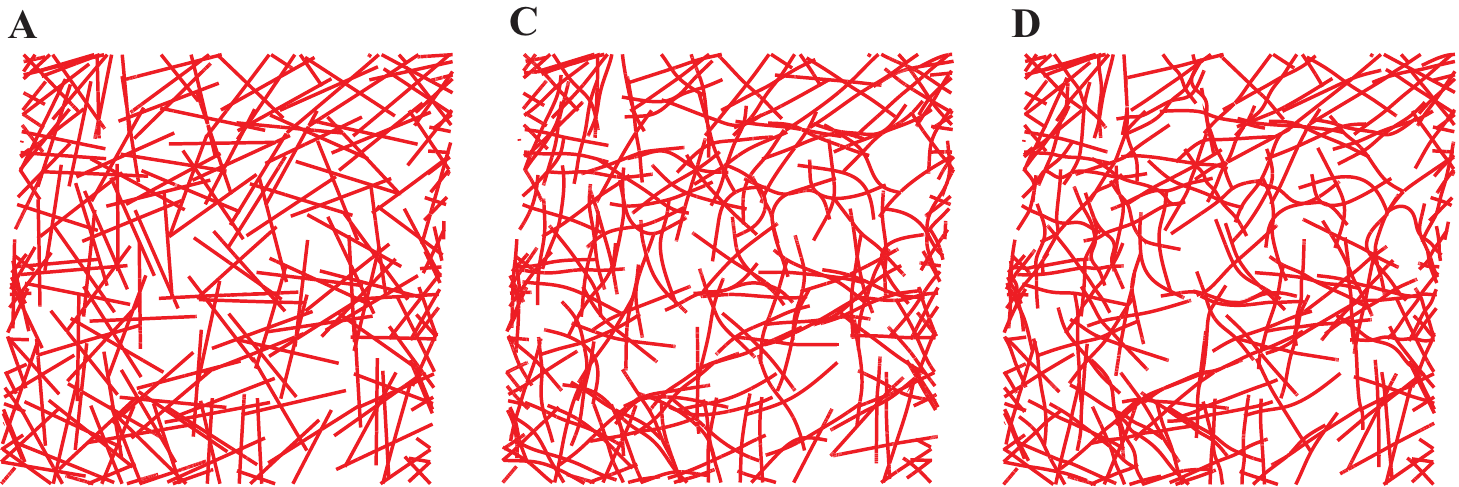}       
  \caption{The deformation of F-actin network corresponding to the points A, C, and D in Fig.~\ref{fig:stiff}.  Deformation is dominated by stretching of compliant crosslinks (A), bending of flexibly crosslinked F-actin (C), or bending of rigidly crosslinked F-actin (D).}
  \label{fig:deformation}
\end{figure}

Next we consider the response of the network with compliant crosslinks in the presence of motors.  At small strains, Fig.~\ref{fig:stiff} shows that  the networks continuously stiffen with increasing density of motors. Even with one motor per filament, the stiffness of the network increases by over one order of magnitude. Networks with about six motors per filament are stiffer than the networks without motors by about two orders of magnitude. To understand the reason for this marked increase in stiffness, we consider a) the ratios of energies associated with different deformation modes (Fig.~\ref{fig:energy}), b) the distributions of the lengths of the crosslinks (Fig.~\ref{fig:clhist}) and c) the shapes of the deformed filaments (Fig.~\ref{fig:deformation}). Fig.~\ref{fig:energy} shows that even at small applied strains, the ratio of the total energy of the crosslinks to the total strain energy of all the filaments decreases significantly with increasing motor density. Indeed, the histogram of crosslink extensional lengths in Fig.~\ref{fig:clhist} confirms that the motors are able to induce stretching of almost all the crosslinks to their contour length, $l_0$, beyond which it is difficult to stretch them. This causes bending of the filaments, which explains the decrease in the ratio of the total energy of stretching of crosslinks to the total energy of the filaments, $E_c/E_f$, with increase in the density of motors. Upon application of a small external load, the deformation is primarily borne by the bending of the filaments leading to stiffer response compared to the case where motors are absent. The dominance of the bending modes is clearly seen in Fig~\ref{fig:deformation}: for small applied strain of 0.05, the filaments in the network without motors are straight whereas significant bending of the filaments can be seen in the network with motors. Also note from Fig.~\ref{fig:clhist} that most of the crosslinks in the former case have not been stretched to the maximum extent. As the applied strain is large ($\sim0.3$), in all cases there is a transition in the deformation modes of the filaments from bending to stretching, at which point the response of all the networks is identical. 

In contrast to the filaments with compliant crosslinks, the stress-strain curves and the incremental moduli of networks with rigid crosslinks are not significantly altered by the presence of motors. Indeed, we find only an increase close to a factor of two in the modulus of the network at small strains. We can understand this by noting that the filaments in rigidly crosslinked networks deform primarily through bending. The motors do not qualitatively change this picture. The effect of the motors on bending deformation is small unlike the large change in the lengths of the crosslinks that they induce in the case of compliant networks. Our calculations therefore provide an explanation for the key role played by the nature of the crosslinks on the mechanical response of active networks. 

It is well known that in uncrosslinked networks the myosin filaments effectively fluidize actin networks by actively sliding antiparallel actin filaments past one another, which can lead to large scale reorganization of the network. However in a crosslinked network,  upon addition of myosin, there was no noticeable change in network structure~\cite{Koenderink09}. When myosin and FLNa were both present, the network still remained homogeneous and unbundled suggesting that relative sliding between the motors and filaments is relatively small, which justifies the treatment of the motors as force dipoles. It is also possible that myosin-driven tension may release FLNa crosslinking. This issue, the effect of potential sliding and perhaps the stretching of myosin motors themselves can be considered in future work using the model we have developed here.

In summary, by using material parameters typical for actin networks with compliant crosslinks, we have shown that myosin II motors generate internal stresses by stretching the crosslinks which in turn pull on the actin filaments. Once the crosslinks are fully stretched to their contour lengths, the differential stiffness of the network can increase by two orders of magnitude in excellent agreement with recent experiments~\cite{Koenderink09}. In addition, our simulations show that motors do not lead to any significant stiffening in the response of rigidly crosslinked networks, also in accord with experiments~\cite{Koenderink09}. These observations underscore the importance of the nature of crosslinks on determining the strain hardening behavior of active networks and provide guidelines for tuning mechanical response of biomimetic systems. 

This work was supported by the US National Science Foundation Grant No. CMMI-0825185. 

\footnotesize{

}


\begin{thebibliography}{17}
\bibitem {Timoshenko} Compared to Euler-Bernoulli beams, this model takes into account shear deformation \cite{Timoshenkobeam}. For the present system, we however found no difference between the two cases.
\end{thebibliography}
\end{document}